\documentclass[a4paper,10pt]{article}
\usepackage[utf8]{inputenc}
\usepackage{graphics}
\usepackage{amssymb}
\usepackage{textcomp}
\usepackage{amsmath}
\usepackage{float}
\usepackage{rotating}
\usepackage{booktabs}
\usepackage{subcaption}
\usepackage{multirow}

\usepackage{array}
\usepackage[usenames, dvipsnames]{color}

\usepackage{mathtools}
\usepackage{pdfpages}
\usepackage{hyperref}
\usepackage{multirow}

\usepackage[left=3cm,right=3cm,top=3cm,bottom=3cm]{geometry}

\usepackage{lineno}
\usepackage{dirtytalk}
\usepackage{float}
\usepackage[square,sort,comma,numbers]{natbib}
\usepackage{hyperref}
\usepackage{graphicx}
\usepackage{physics}
\usepackage{parskip}

\usepackage{multirow}

\begin{document}

\begin{center}
	\textbf{\large Closing the Door on the ``Puzzle of Decoherence'' of Annihilation Quanta}\\
	\vspace*{0.5cm}		
	
	{ Siddharth Parashari$^{a,*}$, Damir Bosnar$^{a}$, Ivica Fri\v{s}\v{c}i\'c$^{a}$, Zdenka Kuncic$^{b}$, Mihael Makek$^{a,*}$}
	\\              
	\vspace*{0.3cm}
	\small
	$^{a}$Department of Physics, Faculty of Science, University of Zagreb, Bijeni\v{c}ka c. 32, 10000 Zagreb, Croatia.\\
	$^{b}$School of Physics, University of Sydney, Sydney, New South Wales, 2006, Australia\\

\end{center}

\vspace*{1cm}
\section*{Abstract}
In para-positronium annihilation, exploration of the polarization correlations of the emerging gamma quanta has gained interest, since it offers a possibility to improve signal-to-background in medical imaging using positron emission tomography. The annihilation quanta, which are predicted to be in an entangled state, have orthogonal polarizations and this property may be exploited to discriminate them from two uncorrelated gamma photons contributing to the background. Recent experimental studies of polarization correlations of the annihilation quanta after a prior Compton scattering of one of them, had rather different conclusions regarding the strength of the correlation after the scattering, showing its puzzling nature. The scattering was described as a decoherence process. In the present work, we perform for the first time, a study of the polarization correlations of annihilation quanta after decoherence via Compton scattering in the angular range $0^\circ - 50^\circ$ using single-layer gamma ray polarimeters.  In addition, we compare the measured polarization correlations after Compton scattering at $30^\circ$ with an active and a passive scatterer element. The results indicate that the correlation, expressed in terms of the polarimetric modulation factor, shows no significant difference at small scattering angles ($0^\circ-30^\circ$) compared to the correlation measured for direct photons, while lower modulation was observed for $50^\circ$ scattering angle.

\vspace*{0.5cm}

{\textbf {Keywords:}} Para-Positronium annihilation ; Quantum entanglement ; Decoherence ; Gamma Polarization ; Positron Emission Tomography 
\vspace*{5.5cm}\\
\begin{flushleft}
	\noindent\rule{4cm}{1pt}\\
	$^{*}$Corresponding authors\\
		Email addresses: siddharth@phy.hr(Siddharth Parashari), 
	makek@phy.hr (Mihael Makek)
\end{flushleft}
\newpage

\newpage
\section{Introduction}
\label{sec:intro}
The correlation of gamma  photons emerging from an annihilating positronium system has recently become an increasingly interesting research topic with a potential to bring substantial innovations to medical imaging with Positron Emission Tomography (PET) \cite{McNamara,Toghyani,Makek_NIM,Watts,SidNIM}. In para-positronium annihilation, two back-to-back $\gamma$ quanta of 511 keV are produced with orthogonal polarizations and are predicted to be in an entangled state. This state can be described by the so called Bell state wave function $\mathrm{\ket{\psi}\!=\!(\ket{X}_-\ket{Y}_+-\ket{Y}_-\ket{X}_+)/\!\sqrt{2}}$, where, $\mathrm{\ket{X}_{+,-}}$ and $\mathrm{\ket{Y}_{+,-}}$ represent a quantum polarized in (X,Y) propagating in (+,-) $\hat{z}$ direction \cite{Bohm,Bell,Snyder,Ward}. The measurement of correlations of annihilation quanta has a long history in physics, starting from the experimental scheme proposed by J. Wheeler \cite{Wheeler} to test the predicted correlation of the polarizations of the annihilation photons, where the azimuthal angle difference ($\mathrm{\Delta\phi}$) between the scattering planes in double Compton scattering process should have maxima and minima at $\mathrm{90^\circ}$ and $\mathrm{0^\circ}$, respectively. Similar behavior is also predicted by Pryce and Ward \cite{Pryce} and Snyder \textit{et al.} \cite{Snyder} by employing the Klein-Nishina approach \cite{K-N}, where the cross-section of a double Compton scattering process is given by \cite{Ward,Pryce},  
\begin{linenomath*}
\begin{equation}
\mathrm{ \frac{d^2\sigma}{d\Omega_1\!d\Omega_2}\!\!=\!\!\frac{r_0^4}{16}\left[F(\theta_1\!)F(\theta_2\!)-\!G(\theta_1\!)G(\theta_2\!)cos(2\Delta\phi)\right] }
\label{DDCS}
\end{equation}
\end{linenomath*}
with $\mathrm{r_0}$ being the classical electron radius, $\mathrm{d\Omega_{1,2}}$ are the solid angles, $\mathrm{\theta_{1,2}}$ are the Compton scattering angles, $\mathrm{\Delta\phi\!=\!(\phi_1-\phi_2})$ is the azimuthal scattering angle difference of gamma particle 1 and 2, respectively. $\mathrm{F(\theta_{1,2})\!=\!\frac{2+(1-cos\theta_{1,2})^3}{(2-cos\theta_{1,2})^3}}$ and $\mathrm{G(\theta_{1,2})\!=\!\frac{sin^2\theta_{1,2}}{(2-cos\theta_{1,2})^2}}$ are kinematic factors. The ratio,
\begin{linenomath*}
\begin{equation}
A(\theta)\!=\!\frac{N(\theta,\phi\!=\!90^\circ)\!-\!N(\theta,\phi\!=\!0^\circ)}{N(\theta,\phi\!=\!90^\circ)\!+\!N(\theta,\phi\!=\!0^\circ)}
\label{analyzingpower}
\end{equation}
\end{linenomath*}
defines the analyzing power of the polarimeter \cite{Knights}, where $\mathrm{N(\theta,\phi)}$ is the number of incoming photons measured  at scattering angle $\theta$ and azimuthal angle $\phi$. The product of the analyzing powers $\mathrm{A(\theta_1)}$ and $\mathrm{A(\theta_2)}$ is known as the modulation factor ($\mu$) and it measures the polarimetric sensitivity of the detection system. Bohm and Aharonov \cite{Bohm} recognized the azimuthal angle correlation in the double Compton scattering of annihilation photons could be considered an example of entanglement discussed by Einstein, Podolsky, and Rosen \cite{EPR,Einstein}. Initially, it was shown theoretically that the ratio, $\mathrm{R\!=\!(1\!+\!\mu)/(1\!-\!\mu)}$ of the scattering probabilities at $\mathrm{\Delta\phi\!=\!90^\circ}$ and $\mathrm{\Delta\phi\!=\!0^\circ}$, for the wave function in the predicted entangled (separable) state is $\mathrm{R\!\approx\!2.8 \; (1.63)}$ \cite{Bohm,Snyder,Ward,Pryce}. 
However, the question of evidence of entanglement in Compton scattering of annihilation quanta has been recently readdressed by theoretical works \cite{Hiesmayr,Caradonna} and it remains an open topic.  The early experiments measured the modulation factors \cite{Wu,Langhoff,Kasday,Faraci,Kasday2,Wilson,Bruno,Bertolini}, among which the most precise measurements were performed by Langhoff \cite{Langhoff} and Kasday \textit{et al.} \cite{Kasday2}, obtaining $\mathrm{R} = \mathrm{2.47\pm0.07}$ and $\mathrm{R} = \mathrm{2.33\pm0.10}$, respectively, which were in good agreement with the predicted value (eq. \ref{DDCS}) considering the finite geometry of the detectors. 

Recently, Watts \textit{et al.} \cite{Watts} used a simple polarimetric setup based on two Cadmium Zinc Telluride matrices to measure the azimuthal angle correlations of annihilation quanta directly from the source and in another configuration with a passive scatterer. In the latter,
a Compton scattering may occur in the path of the photon prior to its detection in the polarimeter. 
In the measurement of the quanta direct from the source, they obtained a clear modulation of the azimuthal distribution with the measured $R\!=\!1.95\pm0.07$ for $\theta_{1,2}\!=\!93^\circ\!-\!103^\circ$. In the configuration with the scatterer positioned at $\mathrm{33^\circ}$ relative to the initial direction, the results indicated the lack of modulation, within the experiment's precision, which they described as a \say{decohering} process. 

In another experiment, Abdurashitov \textit{et al.} \cite{Abdurashitov} used a setup of two gamma-ray polarimeters, each consisting of 16 NaI(Tl) scintillator detectors positioned on a ring, with a plastic scatterer placed at the center. Such geometry enabled precise measurement of the azimuthal angle correlations in events when both annihilation quanta underwent Compton scattering at $\theta_{1,2}=90^\circ$ yielding the modulation of $\mathrm{R\;(\mu) = 2.44\pm0.02}\;(0.418\pm0.003)$. In the modified version of the setup, an active scatterer (GAGG  scintillator) was placed in front of one plastic scintillator to induce decoherence by Compton scattering one of the photons before entering the polarimeter. The scattered photons then underwent another scattering in the polarimeters yielding the modulation of the azimuthal angle difference of $\mathrm{R \;(\mu) =  2.41\pm0.10}\;(0.414\!\pm\!0.017)$, a result compatible with the one obtained without intentional decoherence. 

The results of Abdurashitov \textit{et al.} \cite{Abdurashitov} suggest that the correlation of the azimuthal angles of Compton scattered annihilation quanta is not affected by a prior scattering, at least at small scattering angles $\theta_{scat}\!\approx\!0^\circ$, which disagrees with the finding of Watts \textit{et al.} \cite{Watts} suggesting the loss of correlation at $\theta_{scat}\!\approx\!30^\circ$. The clarification of these findings is important on it own merits, but it is also relevant for the implementation of the polarization measurement in PET, where in-silico studies suggested a potential benefit of discriminating the correlated signal events from the uncorrelated background \cite{Toghyani}. 

To resolve the \say{decoherence puzzle} arising from the previous results, we performed the most comprehensive set of measurements to date. The necessity to resolve this puzzle has also been brought up by Sharma et al., \cite{Sharma}. Our setup based on the single-layer gamma-ray polarimeter concept \cite{Makek_NIM} was able to measure the azimuthal correlation of the annihilation quanta \cite{AMK,SidNIM}. An active scatterer was placed in the path of one annihilation photon to tag the events where that gamma undergoes a prior scattering. One of the detectors could be rotated around the initial direction of the annihilation gamma, enabling the azimuthal correlation measurements at different angles of the prior Compton scattering, $\theta_{scat}$. Thus this setup enabled recreation of the kinematic conditions similar to those in \cite{Abdurashitov} ($\theta_{scat}\approx 0^\circ$) and \cite{Watts} ($\theta_{scat}\approx 30^\circ$), while also significantly extending the explored phase-space to measurements with $\theta_{scat}=10^\circ$ and $\theta_{scat}=50^\circ$.  

\begin{figure}[t]
    \centering
    \includegraphics[scale=.4]{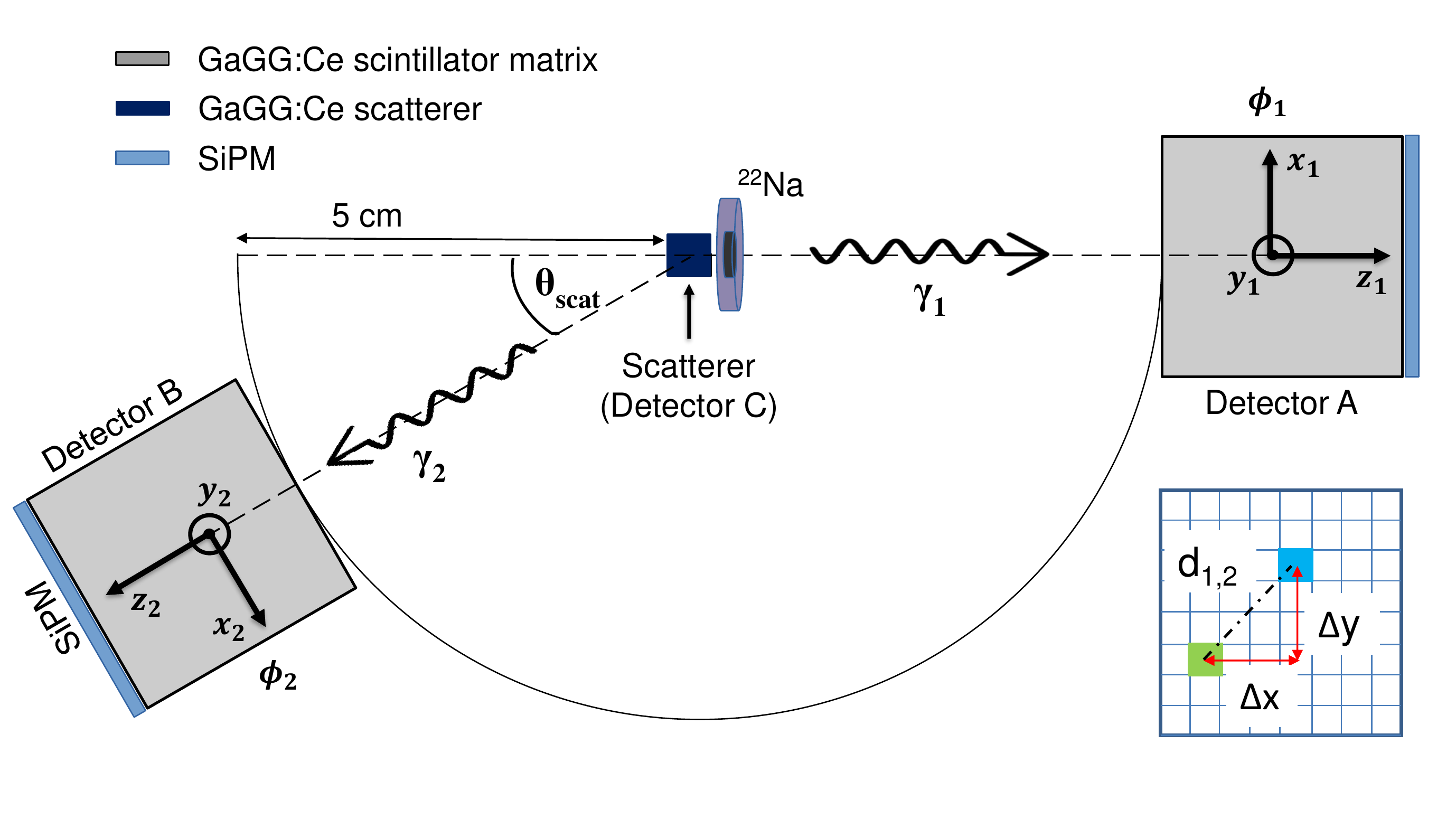}
    \caption{Schematic diagram of the experimental setup (top view). The azimuthal angles $\phi_{1,2}$ (eq. \ref{compton_theta_phi}) and the inter-pixel distances $d_{1,2}$ are deduced from the relative positions of the fired pixels in the respective module.}
    \label{fig:schematics}
\end{figure}
\begin{figure*}[t]
    \centering
    \includegraphics[width=6.5cm,height=6.0cm]{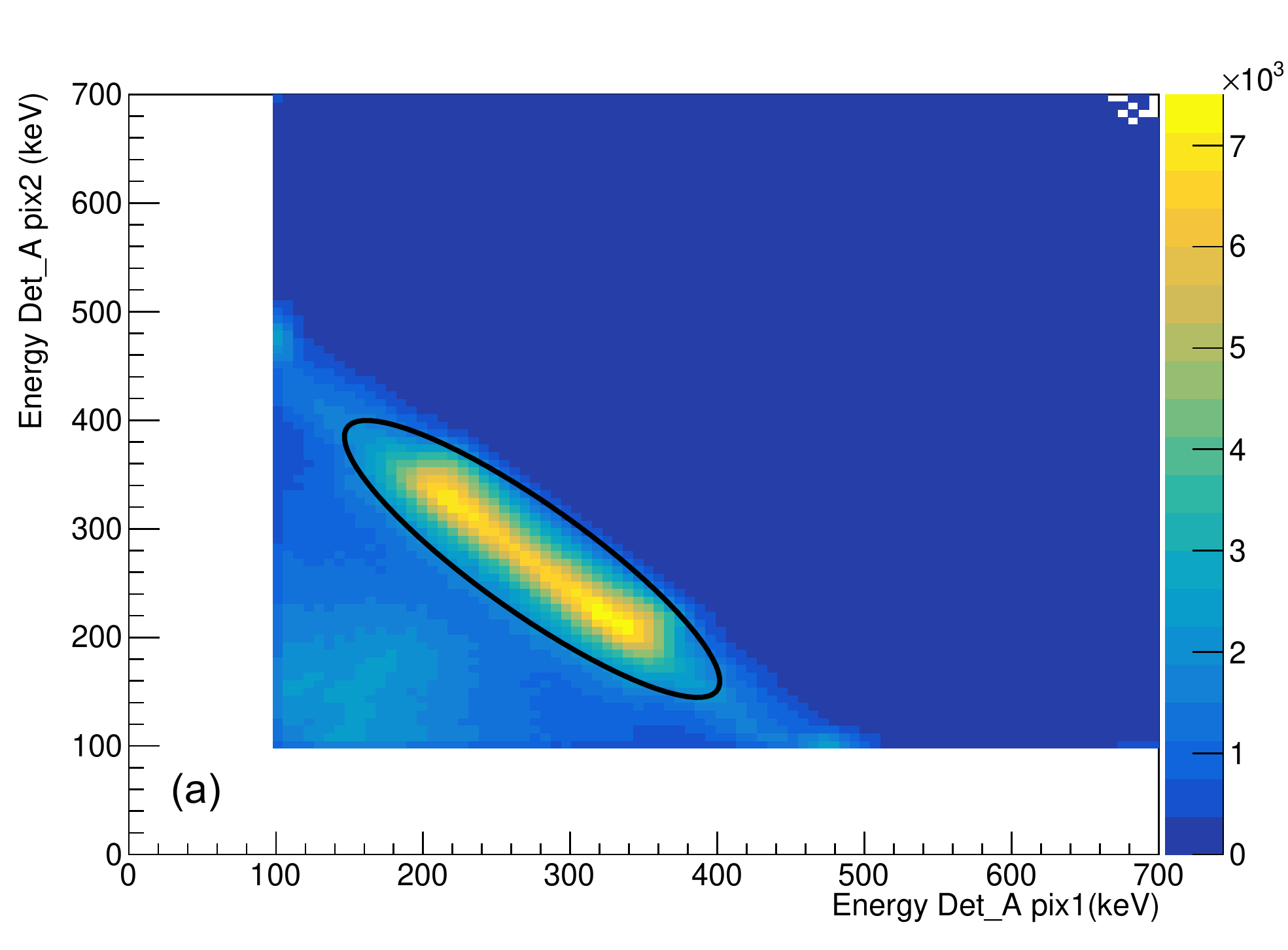}
    \includegraphics[width=6.5cm,height=6.0cm]{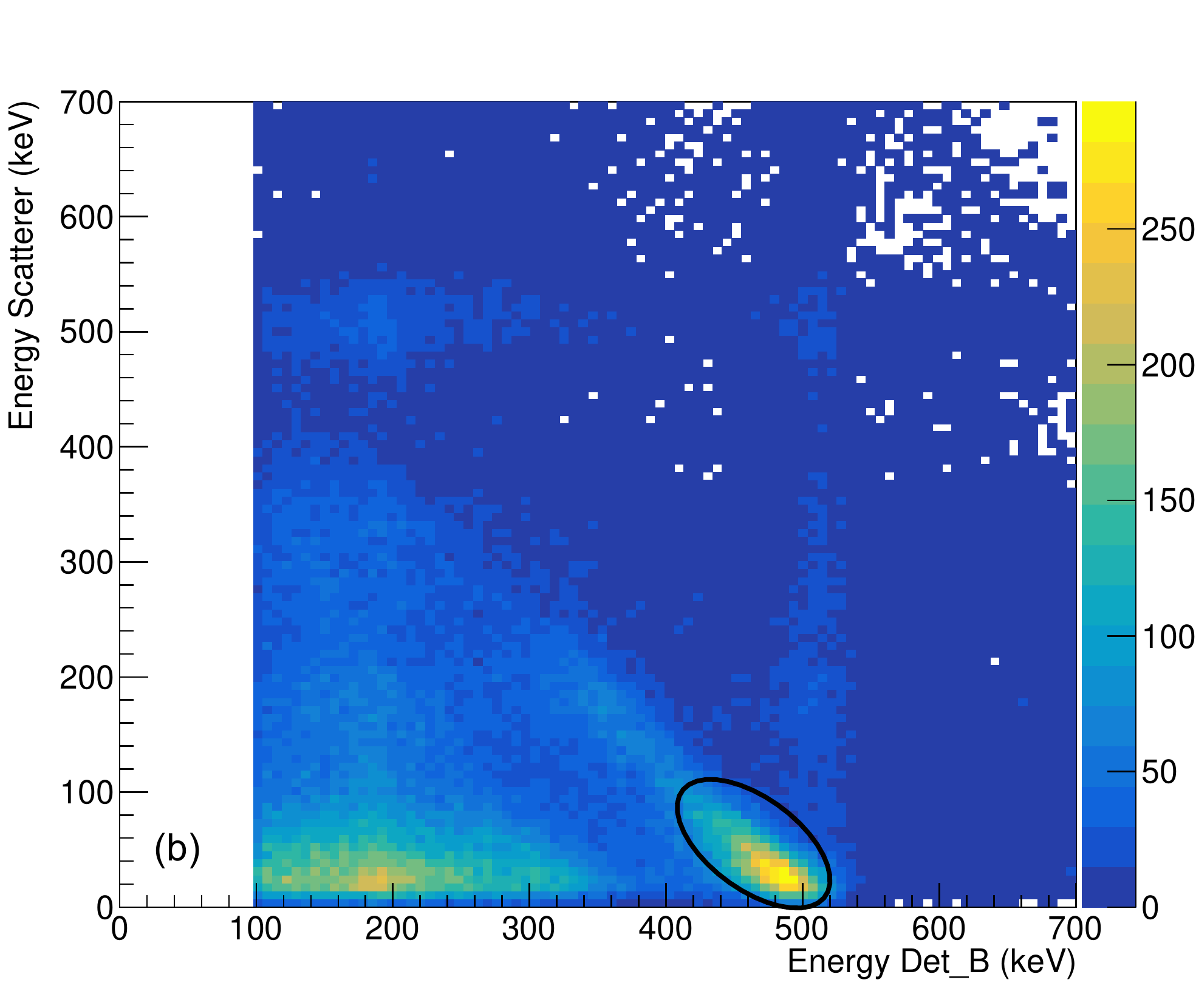}
    \includegraphics[width=6.5cm,height=6.0cm]{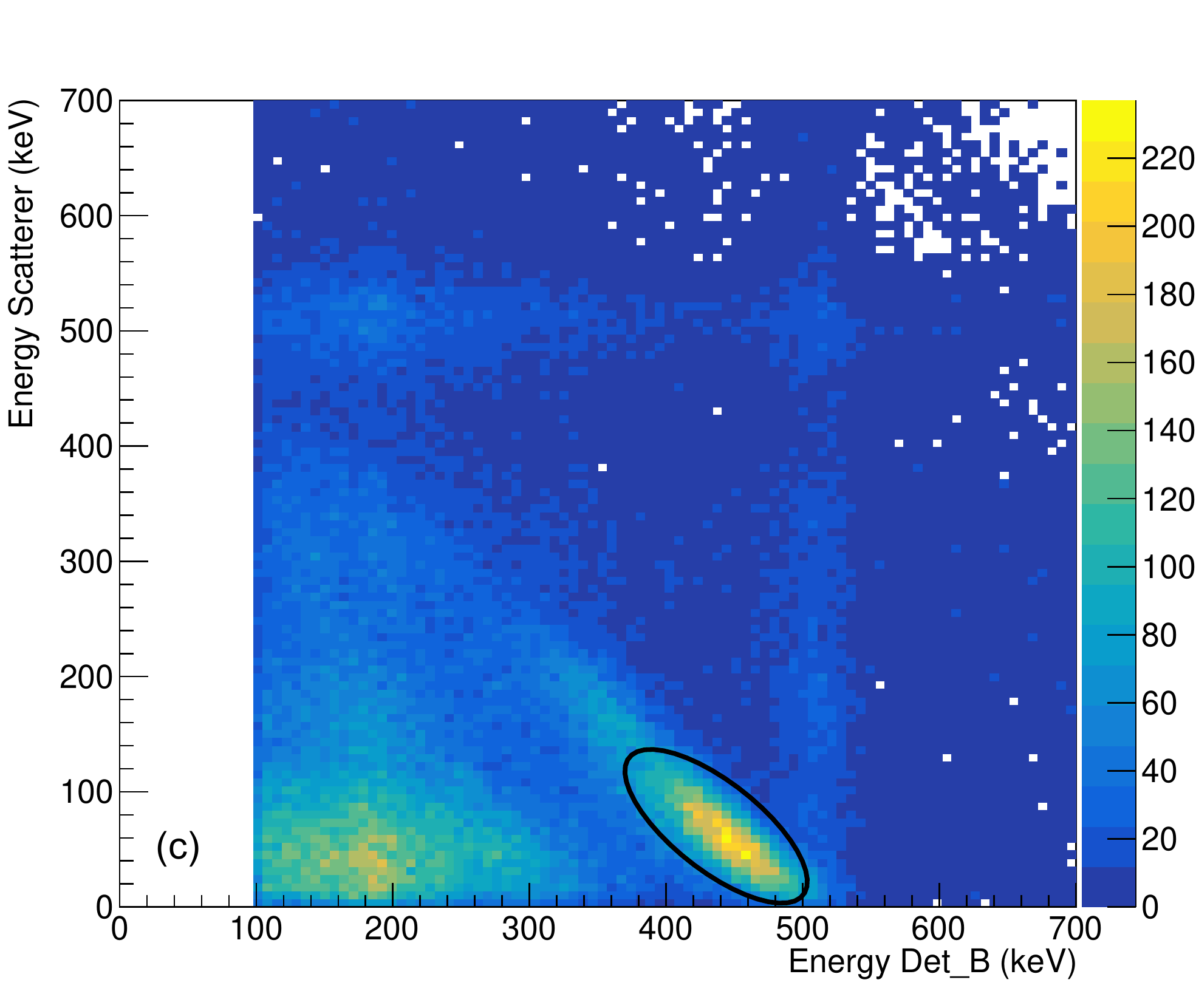}
    \includegraphics[width=6.5cm,height=6.0cm]{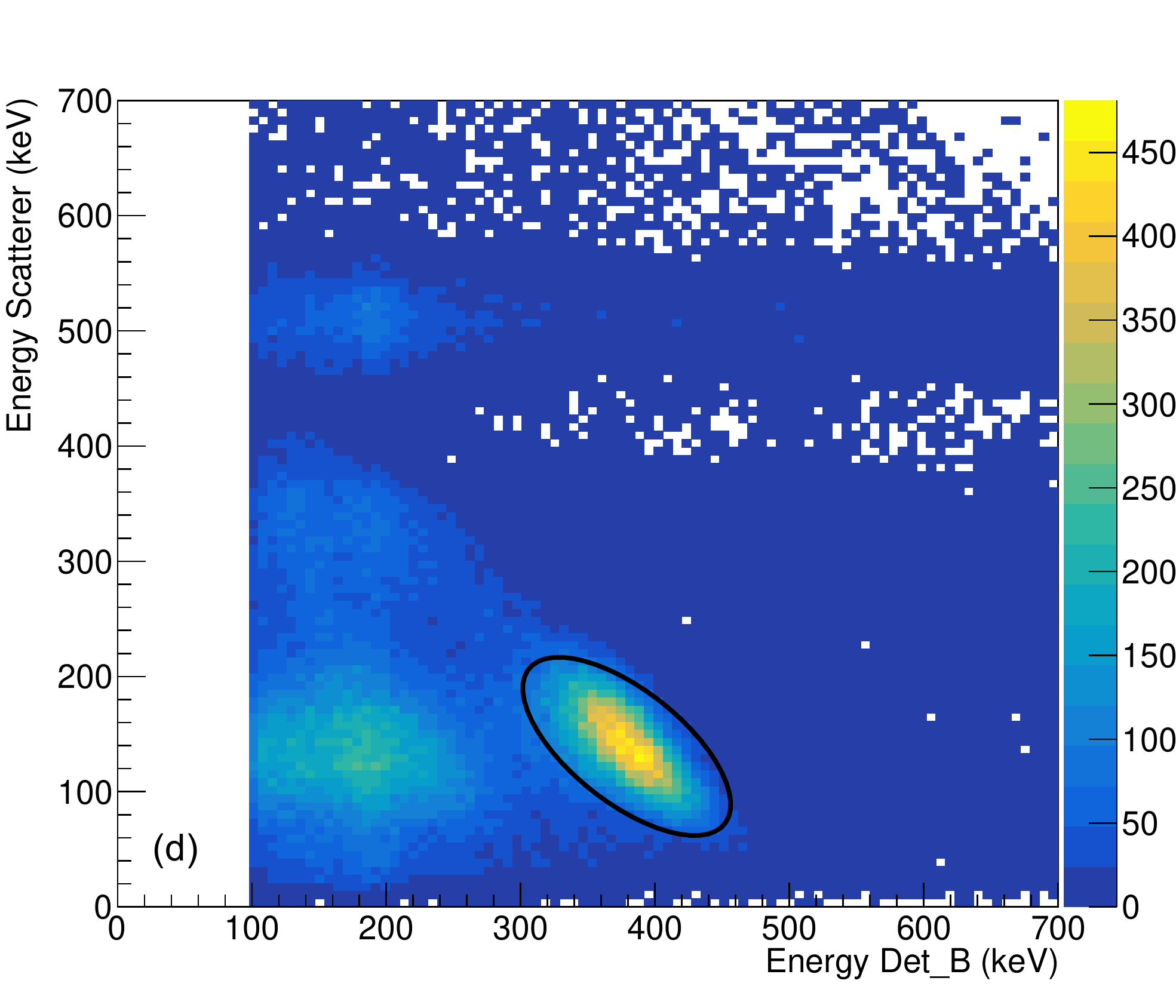}
    \caption{Energy correlation between two pixels fired in a Compton event in Detector A \textbf{(a)}, energy correlation between Detector B and the scatterer Detector C at $\theta_{scat} = 10^\circ$ \textbf{(b)}, at  $\theta_{scat} = 30^\circ$ \textbf{(c)}, and at $\theta_{scat} = 50^\circ$ \textbf{(d)}. The selected events are shown with an ellipse. }
    \label{fig:b_vs_c}
\end{figure*}

\section{Methodology}
\label{sec:exp}
\subsection{Experimental setup}
The experimental setup consists of two single-layer gamma polarimeters \cite{SidNIM}, denoted as Detector A and B and a scatterer scintillator, denoted Detector C as shown in Figure \ref{fig:schematics}. Each polarimeter encompasses $8\times8$ GAGG:Ce scintillator matrix with crystal dimensions $1.9\times1.9\times20 \;\mathrm{mm^3}$ and 2.2 mm pitch. The matrix is read out on one end by a silicon-photomultiplier (SiPM) array, with one-to-one match of crystals and SiPMs. The mean energy resolution (FWHM) of the GAGG:Ce detectors was $\mathrm{8.1\pm0.5\%}$ at $\mathrm{511}$ keV. The scatterer (Detector C) was a single scintillating crystal of GAGG:Ce of $3.0\times3.0\times20 \;\mathrm{mm^3}$ wrapped with teflon. It was read out by one SiPM of a 8x8 SiPM array (KETEK PA3325) and its energy resolution was $12.1 \pm 0.3\%$ at 511 keV.   
The experiment was performed in the temperature controlled environment keeping the temperature of the Detectors A and B at $18\pm1\;^\circ C$. The temperature of the scatterer was further reduced to $15\pm1^\circ C$ by a Peltier-based cooling system, to improve the sensitivity for low energy events. 
The data were acquired using the data acquisition and processing system TOFPET2 \cite{tofpet1, tofpet2}. A modified set of ASIC parameters together with a lower value of time and energy thresholds were used for the data acquisition to enable the acquisition of events with low energy deposits. To do so, the trigger threshold parameters were tuned globally to lower the baseline for dark count rejection and enable trigger for low energy events. 
The energy measured in each pixel was corrected for SiPM saturation and calibrated using the 511 keV photo-peak from the direct $\gamma$-rays. The calibration was independently checked with the 32 keV and 662 keV peaks of $\mathrm{^{137}Cs}$ and was found to be consistent to the level of $\leq1\%$.

In the experiment, Detector A was detecting the $\mathrm{\gamma\!-\!ray}$ coming directly from the annihilation event (denoted $\gamma_1$), while Detector B was detecting the $\mathrm{\gamma_2}$ after scattering in the Detector C (the scatterer), which was introduced to induce decoherence by Compton scattering the annihilation photon and to tag such events. Detectors A and B were kept at a fixed distance of 5 cm from the scatterer. A $\mathrm{^{22}Na}$-source (1 mm diam., activity $\!\approx$ 370 kBq) was placed 1 cm from the scatterer between the scatterer and Detector A. The angular coverage of Detectors A and B was $\pm10.1^\circ$ in this geometry.
 The events were recorded with Detector B placed at different mean scattering angles, $\mathrm{\theta_{scat}}$, of $\mathrm{0^\circ}$, $\mathrm{10^\circ}$, $\mathrm{30^\circ}$, and $\mathrm{50^\circ}$, while the Detector A was fixed.
 An additional measurement was performed to recreate the conditions of \cite{Watts} with Detector B at $30^\circ$, in which the scatterer was made passive simply by switching it off, so it would not contribute to trigger or data acquisition.
\begin{figure}[t]
    \centering
    \includegraphics[width=7cm,height=7cm]{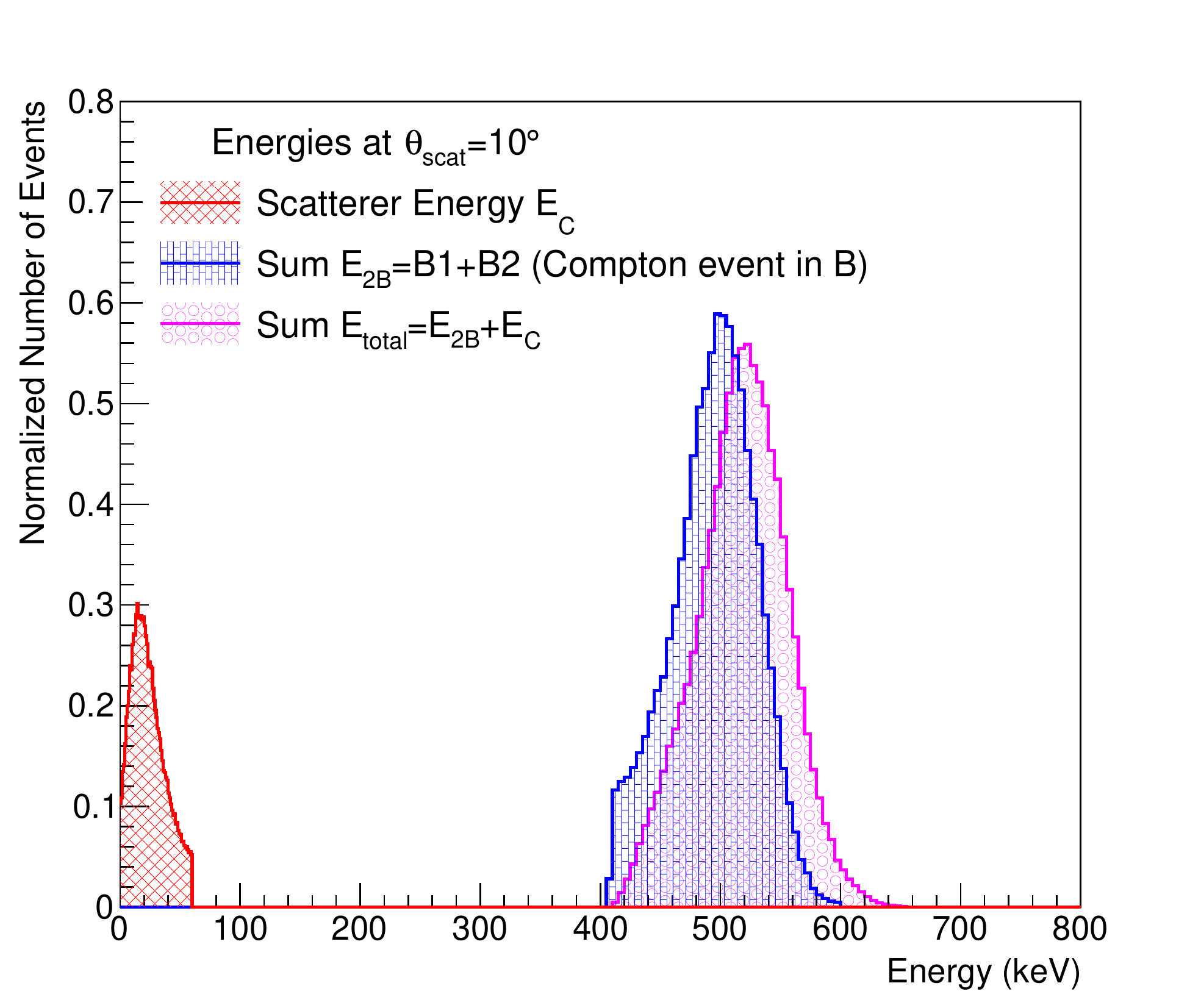}
    \caption{Energies shared among scatterer and two pixels of Detector B for $\theta_{scat} = 10^\circ$. 
    The total sum of energies adds up to 511 keV.}
    \label{fig:energies}
\end{figure}


\begin{figure}[ht]
    \centering
    \includegraphics[width=7cm,height=7cm]{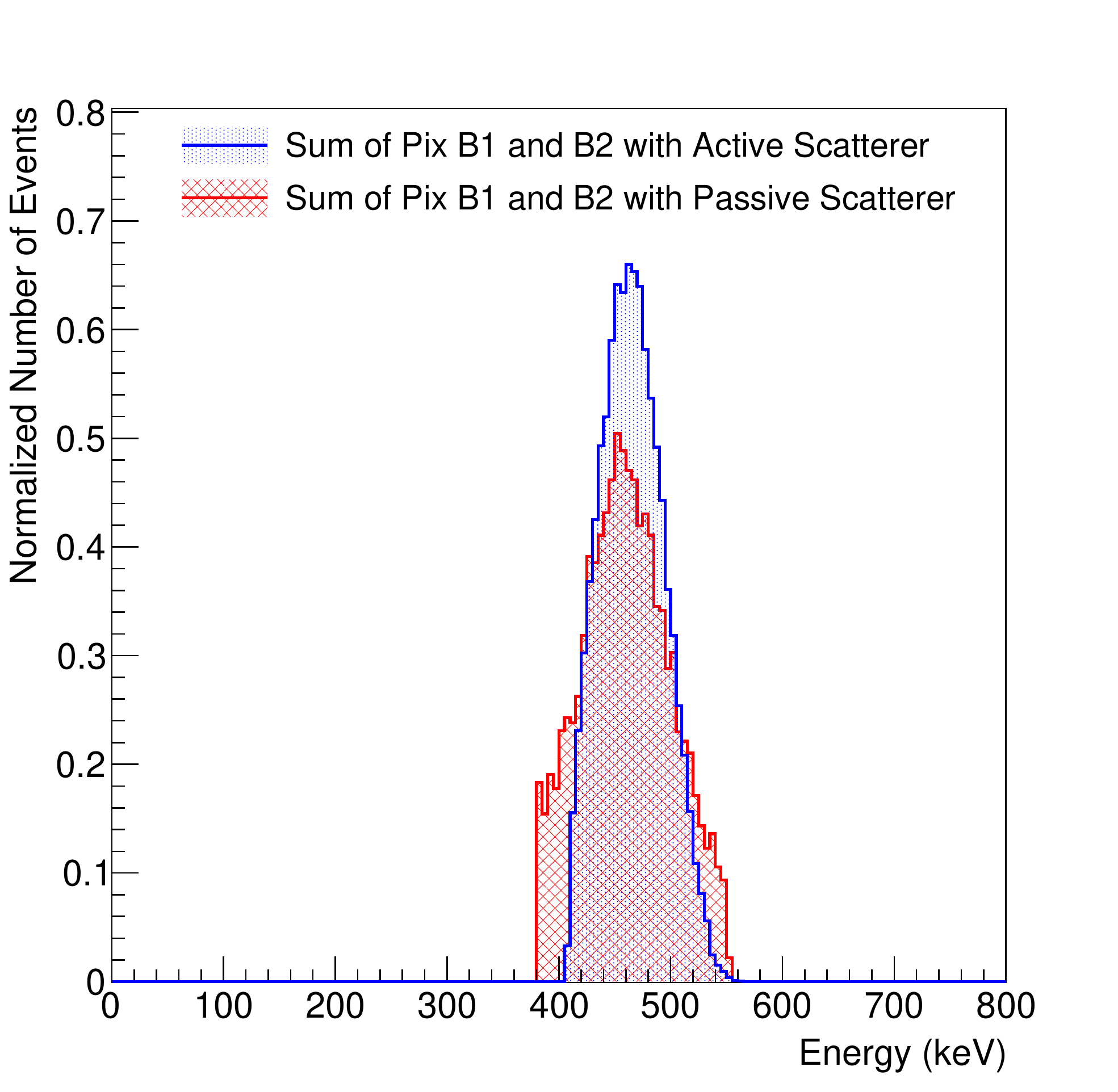}
    \caption{The sum of energies of two pixels fired in Detector B, for the selected Compton events at $\theta_{scat} = 30^\circ$, active (blue) and passive scatterer (red), using their respective selection criteria.}
    \label{fig:Energy_active_vs_passive}
\end{figure}
\subsection{Event selection}
\label{sec:event_selec}
Data acquisition was triggered by coincidence events in Detector A and either of Detectors B or C ($\mathrm{A \cap (B \cup C)}$).  
Further event selection was done in the analysis. To select the Compton events that occur in Detectors A or B we required the event to have a multiplicity of 2 fired pixels in the module, where a lower bound of 100 keV was applied to count a pixel as fired to avoid possible noise and cross-talk events. For Detector A, the events where two pixels fired were additionally filtered by requiring the sum of energies of the two fired pixels within $511\pm70$ keV, which corresponds to $3\sigma$ around the peak value, and to obey the Compton scattering kinematics. The resulting selection is shown by the outlined region in Figure \ref{fig:b_vs_c} (a). 
To select the annihilation photons that underwent Compton scattering in Detector C and a subsequent Compton scattering following the full absorption in Detector B we 
required the sum of energies of two fired pixels in Detector B with Detector C are within $511\pm3\sigma$ keV. However different types of scattering events can satisfy this condition, which is why we additionally filtered those events that obey the Compton kinematics for scattering at a given $\theta_{scat}$. The correlation of energies between Detector C and Detector B under different scattering angles $(\theta_{scat})$ is shown in Figure \ref{fig:b_vs_c} (b)-(d), which clearly demonstrates that the energy deposition in Detector C increases with the increasing $\theta_{scat}$, as expected according to Compton kinematics. An example of the energy share of $\gamma_{2}$ in pixels selected for $\theta_{scat}\!=\!10^\circ$ is shown in Figure \ref{fig:energies}.

A correlation-baseline measurement (without intentional decoherence), at  $\theta_{scat}=0^\circ$, was performed by selecting events where Detector C did not fire and both energy deposits in Detectors A and B were within $511\pm3\sigma$ keV. 

An additional measurement was done with the passive scatterer, where the bias voltage of Detector C was switched off. In that case, event selection was done  solely based on data from Detectors A and B. The angle of Detector B  was set to $\theta_{scat}\!\!=\!\!30^\circ$ and its the distance from the scatterer was increased to 7.5 cm to avoid direct coincidences between Detectors A and B. In this case, Detector B had an angular coverage of $\pm6.8^\circ$. 
Although direct coincidences were avoided by the setup geometry, random coincidences of two annihilation photons from different events could contribute to the expected kinematic region. To avoid such unwanted events we additionally selected true coincidences based on their coincidence time. Hence, the events in which triggering time difference of the corresponding pixels in Detectors A and B was, $\Delta t=\abs{t_1-t_2} < 1.95$ ns, were selected corresponding to $\pm 3 \sigma$ cut on the coincidence time peak. The energy of the selected Compton events in Detector B for $\mathrm{\theta_{scat}\!\!=\!\!30^\circ}$ is shown in Figure \ref{fig:Energy_active_vs_passive} and is consistent with the energy spectrum of such events obtained with the active scatterer.   



\begin{figure*}[t]
    \centering
     \includegraphics[scale=0.669]{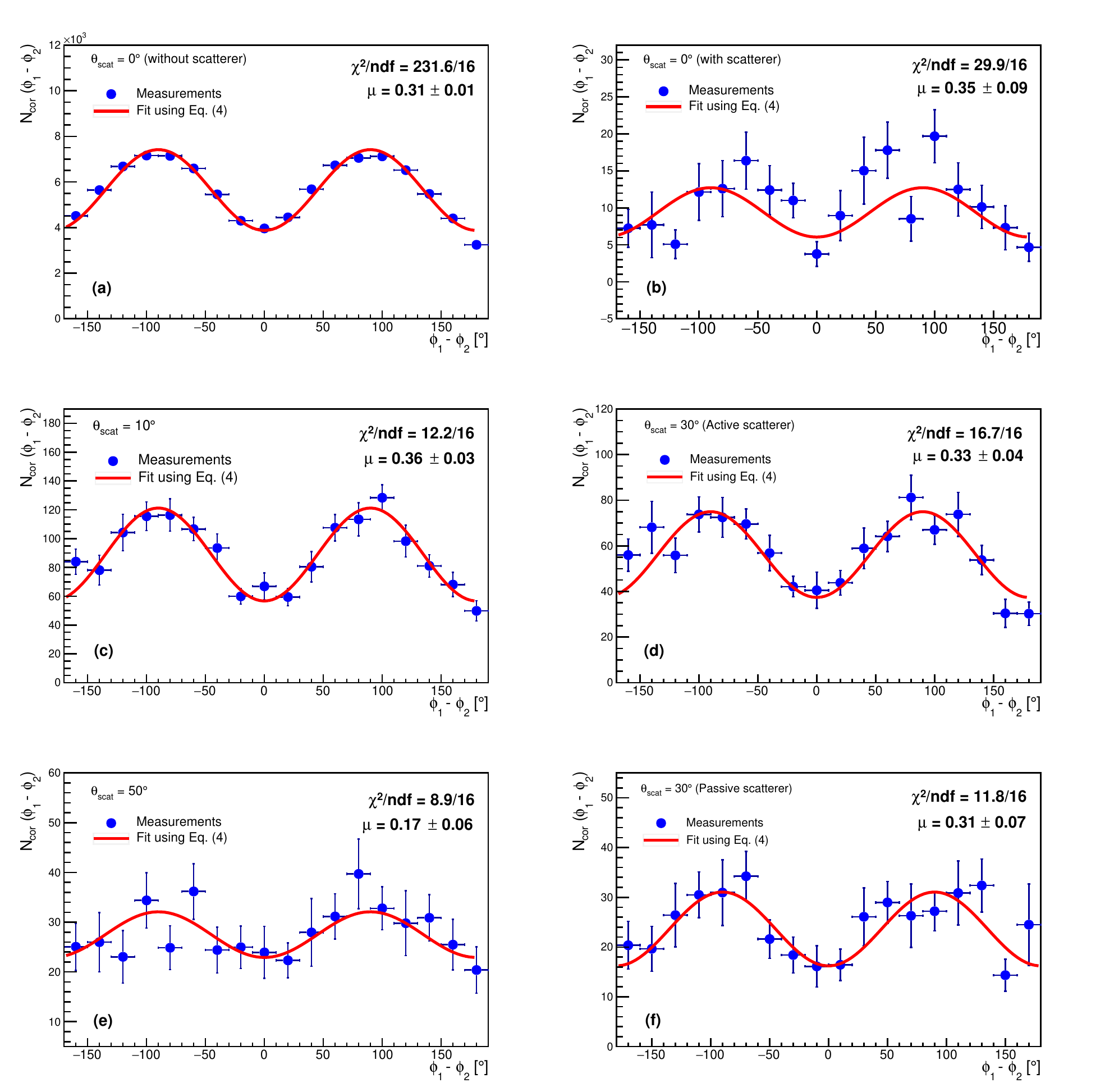} 
    \caption{Azimuthal angle difference distributions for \textbf{(a)} direct gammas from the source, with active scatterer \textbf{(b)} $\mathrm{\theta_{scat}\!=\!0^\circ}$ , \textbf{(c)} $\mathrm{\theta_{scat}\!=\!10^\circ}$, \textbf{(d)} $\mathrm{\theta_{scat}\!=\!30^\circ}$, \textbf{(e)} $\mathrm{\theta_{scat}\!=\!50^\circ}$, and \textbf{(f)} $\mathrm{\theta_{scat}\!=\!30^\circ}$ with passive scatterer, respectively. 
    The range of Compton scattering angles, $\mathrm{\theta_{1,2}}$, in each plot was selected to target maximum expected modulation (see section \ref{sec:res}). 
    Only the statistical uncertainties are shown.}
    \label{fig:polcor}
\end{figure*}

\subsection{Determination of azimuthal correlations}
\label{sec:recon_angles}
For the events where both gammas underwent Compton scattering in Detectors A and B according to the conditions above, we deduce the Compton scattering angle ($\theta$) and the azimuthal angle ($\phi$) in each module as:
\begin{linenomath*}
\begin{equation}
\mathrm{ 
\theta\!=\! acos\left(\frac{m_ec^2}{E_{px_1}\!\!+\!E_{px_2}}\!-\!\frac{m_ec^2}{E_{{px}_2}}\!+\!1\right)\!;}\;   
\mathrm{\phi\!=\!atan\left(\frac{\Delta y}{\Delta x}\right)}
 \label{compton_theta_phi}
\end{equation}
\end{linenomath*}
Due to the ambiguity in the determination of the first and second pixels fired in Compton scattering, by the recoil electron and the scattered gamma, respectively, we have to assume that the first interaction (absorption of the recoil electron) occurs in the pixel with the lower energy deposit  ($\mathrm{E_{px_{1}}\!=\!E_{e'}}$, $\mathrm{E_{px_{2}}\!=\!E_{\gamma'}}$) since the cross-section and the detector configuration favor forward scattering. According to simulations, for 511 keV gammas scattering at angles $70^\circ<\theta_{{1,2}}<90^\circ$, this is true in approximately $52\%$ of events\cite{AMK}. This ambiguity does not play a significant role in determination of $\Delta\phi=\phi_1-\phi_2$ since it is an angle between the two scattering planes. The $\phi$ angle is reconstructed from $\Delta x$ and $\Delta y$, the distances of the fired pixel centers in the plane perpendicular to the longer crystal axis. 



For the events that satisfy the Compton selection criteria and for a given range of the reconstructed angles $\theta_{1,2}$, we obtained the distribution of the azimuthal angle differences, $\mathrm{N(\phi_1-\phi_2)}$, where $\mathrm{\phi_{1,2}}$ are the azimuthal angles of the Compton events in Detector A and B, respectively. The $\mathrm{N(\phi_1-\phi_2)}$ distributions were then corrected for detector acceptance as: $\mathrm{{N_{cor}(\phi_1-\phi_2)}}=\mathrm{N(\phi_1-\phi_2)}/\mathrm{{N_{mixed}(\phi_1-\phi_2)}}$. The $N_{mixed}$ is the acceptance determined by event-mixing technique \cite{Makek_NIM,SidNIM}, where $\mathrm{(\phi_1-\phi_2)}$ is obtained by taking $\phi_1$ and $\phi_2$ from different randomly chosen events. For each selected Compton event in Detector A, $10^2$ random uncorrelated events are sampled in Detector B. 

The modulation factor, $\mu$, is determined by fitting the acceptance-corrected distribution, $\mathrm{N_{cor}(\phi_1-\phi_2)}$, with
\begin{linenomath*}
\begin{equation}
  \mathrm{N_{cor}(\phi_1-\phi_2)=M[1-\mu \; cos(2(\phi_1-\phi_2))]}
  \label{fitfunction}
\end{equation}
\end{linenomath*}
where $\mathrm{M}$ corresponds to the average amplitude of the distribution. 

The systematic uncertainty of the determined modulation factor reflects the uncertainty in the determination of $\theta_{1,2}$ angles, resulting from the finite energy resolution of the pixels. We found this contribution to the systematic uncertainty to be $\mathrm{5.8\%}$ following the uncertainty of $\sigma_\theta=\!6.5^\circ$. The $\sigma_\theta$ broadens the nominally selected $\theta_{1,2}$ window by $\theta_{1,2}\!\pm\!\sigma_{\theta_{1,2}}$ and underestimates $\mu$ around the maximum value that is achieved for $\theta_{1,2}=82^\circ$\cite{SidNIM}. 

\section{Results and discussion}
\label{sec:res}
The modulation of the azimuthal angle difference was measured for the annihilation quanta emerging from para-positronium detected directly and in cases where $\gamma_2$ was made decoherent following Compton scattering in the scatterer at $\mathrm{\theta_{scat}\!=\!0^\circ}$, $\mathrm{10^\circ}$, $\mathrm{30^\circ}$, and $\mathrm{50^\circ}$. For quanta of 511 keV, the maximum modulation is expected for  $\theta_{1,2}\!\approx\!82^\circ$\cite{Snyder,Ward,Pryce}, therefore, for $\theta_{scat}=0^\circ$, we selected the angular range of $72^\circ<\theta_{1,2}<90^\circ$.
For scattering at $\mathrm{\theta_{scat}\!=\!30^\circ}$ and $\mathrm{\theta_{scat}\!=\!50^\circ}$ the selected angular range in detector B was
 $\mathrm{73^\circ<\theta_{2}<90^\circ}$ and  $\mathrm{74^\circ<\theta_{2}<90^\circ}$, respectively. This is because the analyzing power $\mathrm{A(\theta)\!=\!sin^2\theta/(\epsilon+1/\epsilon-sin^2\theta)}$ (derived from eq. \ref{analyzingpower}) \cite{Knights}, depends on incident gamma-ray energy, since $\mathrm{\epsilon}$ is the ratio of scattered and incident photon energies. Therefore the maximum analyzing power for $\mathrm{E_2}=450$ keV (after scattering at $\mathrm{\theta_{scat}\!=\!30^\circ}$) is achieved at $\mathrm{\theta_2}\approx83^\circ$ and the maximum analyzing power for $\mathrm{E_2}=376$ keV (after scattering at $\mathrm{\theta_{scat}\!=\!50^\circ}$) is achieved at $\mathrm{ \theta_2}\approx84^\circ$.

The acceptance-corrected distributions of the azimuthal angle differences, $\mathrm{N_{cor}(\phi_1-\phi_2)}$, for $\mathrm{\theta_{scat}\!=\!0^\circ}$, $\mathrm{10^\circ}$, $\mathrm{30^\circ}$, and $\mathrm{50^\circ}$ are shown in Figure \ref{fig:polcor}. The distribution for the direct measurement (Figure \ref{fig:polcor} (a)) exhibits the expected behavior, with the maxima at $\pm90^\circ$ indicating the initial orthogonality in the polarizations of the annihilation $\gamma$s. Similar behavior can be observed in the distributions obtained after Compton scattering by an active scatter (Figure \ref{fig:polcor} (b)-(e)) and the passive scatterer (Figure \ref{fig:polcor} (f)).
The extracted  modulation factors are listed in Table 1. They suggest that the strength of the modulation is largely preserved at all measured scattering angles $\mathrm{\theta_{scat}}$, within the precision of the experiment. Such a conclusion seems to be in line with the recent findings of Abdurashitov \textit{et al.} \cite{Abdurashitov} for $\mathrm{\theta_{scat}=0^\circ}$, but additionally extends the examined range up to $\mathrm{\theta_{scat}=30^\circ}$. Moreover, the azimuthal angle modulation is observed at $\mathrm{\theta_{scat}\!=\!30^\circ}$ with a passive scatterer, even though it was not evident from the measurements by Watts \textit{et al.} \cite{Watts}. We observed somewhat lower modulation for $\mathrm{\theta_{scat}\!=\!50^\circ}$, yielding $\mathrm{\mu=0.17\pm0.06}$, which may be a hint of a partial depolarization of $\gamma_{2}$, expected for larger scattering angles \cite{Depaola}, although this result is limited by the statistical precision. 
 
\begin{table}[ht]
    \label{table}
\begin{center}
    \caption{Polarimetric modulation factor, $\mu$, for all measured configurations
    The values of $\mu$ presented are subject to additional systematic underestimation of 5.8\%.
    The data acquisition time $T_{acq.}$ in each setup is listed.}
    \vspace{2mm}
    \begin{tabular}{c c c}
       \hline
        $\theta_{scat}^\circ$ & $\mu\pm\Delta\mu_{stat.}$ & $T_{acq.} (hour)$ \\
        \hline
        \multirow{2}{*}{$0^\circ$}&  $0.31\pm 0.01$  & \multirow{2}{*}{$48^*$}\\
          &$0.35\pm 0.09^*$   & \\
        $10^\circ$&  $0.36\pm 0.03$   & 256\\
        $30^\circ{}^\dag$&  $0.33\pm 0.04$   &205\\
        $30^\circ{}^\ddag$&  $0.31\pm 0.07$   &92\\
        $50^\circ$&  $0.17\pm 0.06$   &90\\
        \hline
    \end{tabular}\\
\end{center}
 $^*${\small
The total data acquisition duration for the direct gammas and the $\theta_{scat}\!=\!0^\circ$ configuration. Following event selection the latter accounts for $\!\approx\!0.15\%$ of the events. 
$^\dag$ $\theta_{scat}\!=\!30^\circ$ with the active scatterer, 
$^\ddag$ $\theta_{scat}\!=\!30^\circ$ with the passive scatterer }
\end{table}


\section{Conclusions}
\label{sec:con}
We measured the azimuthal correlations of the gamma quanta emerging from para-positronium annihilation, based on their Compton scattering in the single-layer gamma-ray polarimeters. Measurements were performed with one of the quanta undergoing a Compton scattering as a decohering process prior to entering the polarimeter and they are compared with the baseline measurement where the annihilation quanta are detected directly from the source. The obtained distributions of the azimuthal angle differences exhibit the modulation with maxima at $\phi_1-\phi_2=\pm90^\circ$ as expected. The results show that the strength of the correlation reflected in the measured polarimetric modulation factor does not significantly differ for the case of direct quanta or the case of decoherent quanta for $\theta_{scat}\!=\!0^\circ$, $10^\circ$, and $30^\circ$. 
This conclusion differs from indications obtained by Watts \textit{et al.} \cite{Watts} for $\theta_{scat}\approx30^\circ$, but it is in line with the results of Abdurashitov \textit{et al.} \cite{Abdurashitov} for $\theta_{scat}\approx0^\circ$.
For $\theta_{scat}\!=\!50^\circ$, we observe lower modulation than at the other scattering angles investigated.
Further statistically more significant measurements, especially at large scattering angles may give a more complete insight into the puzzling nature of the correlations of annihilation quanta.

\section*{Acknowledgements}
This work was supported in part by the “Research Cooperability” Program of the Croatian Science Foundation, funded by the European Union from the European Social Fund under the Operational Programme Efficient Human Resources 2014–2020, grant number PZS-2019-02-5829, in part by QuantiXLie Center of Excellence, a project co-financed by the Croatian Government and European Union through the European Regional Development Fund — the Competitiveness and Cohesion Operational Programme, grant No. KK.01.1.1.01.0004 and in part by EU Horizon 2020 research and innovation programme under project OPSVIO, Grant Agreement No. 101038099.

\section*{Declaration of competing interest}
The authors declare that they have no known competing financial interests or personal relationships that could have appeared to influence the work reported in this paper.

\section*{Data availability}
Data can be made available on request.



\bibliographystyle{unsrt}

\end{document}